\documentclass{ws-procs9x6}

\begin{document}

\def\endpage{\vfill \eject}
\def\leaderfill{\leaders \hbox to 1em{\hss . \hss}\hfill}
\def\index#1#2{\line{ #1 \leaderfill #2}}
\def\etal{et al.}
\def\ie{i.e.}   
\def\ni{$^{56}Ni$}
\def\co{$^{56}Co$}
\def\endpage{\vfill \eject}
\def\leaderfill{\leaders \hbox to 1em{\hss . \hss}\hfill}
\def\index#1#2{\line{ #1 \leaderfill #2}}
\def\ie{i.e.}   
\def\ctr{\centerline}
\def \lta {\mathrel{\vcenter
     {\hbox{$<$}\nointerlineskip\hbox{$\sim$}}}} 
\def \gta {\mathrel{\vcenter
     {\hbox{$>$}\nointerlineskip\hbox{$\sim$}}}} 
\def \m {M$_\odot$}          
\def \mm {M_\odot}           
\def \l {L$_\odot$}          
\def \ll {L_\odot}           
\def \etal {{\it et~al.}}    
\def \degrees{$^\circ$}
\def \ibid#1 {\hbox{\vrule height2.5pt depth-2pt width1.5cm . #1}}
\def \oneskip{\vskip\baselineskip} 
\def \Teff{{\it T}\lower.5ex\hbox{\rm eff}}   
%
\def\gcc{{\rm g cm$^{-3}$}}
\def\gcm3{g~cm$^{-3}$}
\def\g-s{g~s$^{-1}$}
\def\cm3s{cm$^3$~s$^{-1}$}
\def\cms{cm~s$^{-1}$}
\def\kms{km~s$^{-1}$}
\def\erg-s{erg~s$^{-1}$}
\def\degree{$^{\rm o}$}
\def\beq{\begin{equation}}
\def\eeq{\end{equation}}
\def\gr{$\gamma$-ray}
\def\grs{$\gamma$-rays}
\def\grb{GRB}
\def\grbs{GRBs}
\def\apj{ApJ}
\def\apjl{ApJ Lett.}
\def\aap{A\&A}

\title{
Magnetic Fields in Core Collapse Supernovae: Possibilities and Gaps
}

\author{J. Craig Wheeler and Shizuka Akiyama}

\address{ Department of Astronomy University of Texas \\
E-mail: wheel@astro.as.utexas.edu; shizuka@astro.as.utexas.edu}

\maketitle

\abstracts{
Spectropolarimetry of core collapse supernovae has shown that they are
asymmetric and often, but not universally, bi-polar. The Type IIb SN~1993J and 
similar events showed large scatter in the Stokes parameter plane. SN~2002ap 
which showed very high photospheric velocities in early phases revealed that the 
dominant axes associated with hydrogen, with oxygen, and with calcium were all 
oriented substantially differently. Observational programs 
clearly have much more to teach us about the complexity of asymmetric supernovae 
and the physics involved in the asymmetry. Jet-induced supernova models give a 
typical jet/torus structure that is reminiscent of some objects like the Crab 
nebula, SN~1987A and perhaps Cas A. Jets, in turn, may arise from the intrinsic 
rotation and magnetic fields that are expected to accompany core collapse.  We 
summarize the potential importance of the magneto-rotational instability for the 
core collapse problem and sketch some of the effects that large magnetic fields, 
$\sim 10^{15}$ G, may have on the physics of the supernova explosion. Open issues 
in the problem of multi-dimensional magnetic core collapse are summarized and a 
critique is given of some recent MHD collapse calculations. A crucial piece of 
information that can inform the discussion of potential MHD effects even in the 
absence of the explicit inclusion of magnetic fields is to give sufficient 
information from a rotating collapse to at least crudely estimate the 
time-dependent saturation field according to the prescription 
$v_a \sim r \Omega$.  Many studies of rotating collapse produce 
such information, but fail to present it explicitly. 
}


\section{Introduction}

Spectropolarimetry of supernovae has opened up a new window on these
spectacular events and yielded remarkable new insights. A few 
rare, nearby supernovae and supernova remnants have revealed asymmetric images.
Among these are the Crab nebula with its prominent jet/torus structure revealed by 
CXO, SN~1987A (Wang et al. 2002a) and Cas A (Fesen 2001; Hwang et al. 2004).
It was not clear on this limited basis whether or not the strong asymmetries of 
these objects was important to the intrinsic process of the explosion. 
Spectropolarimetry has extended our knowledge of the composition-dependent 
geometry of core-collapse supernovae to numerous extragalactic supernovae. 
Spectropolarimetry of supernovae probes the geometrical structure of matter
shed by a star before it explodes and the structure of the ejecta of the explosion 
with an effective spatial resolution far superior to any envisaged optical
interferometry (Wang et al. 2002b). The structure revealed is closely related to 
the explosion mechanisms and to the progenitor systems. 

Spectropolarimetry of supernovae continues to show
that all core-collapse supernovae (those associated with young
populations; Type II, Type Ib/c) are polarized and hence
substantially asymmetric (Wang et al. 1996; Wang et al. 2001, 2002a,b, 
2003a,b; Leonard et al. 2000; Leonard \& Filippenko 2001; 
Leonard et al. 2001, 2002). 
The understanding that core-collapse supernovae are routinely 
asymmetric developed in parallel with the discovery that gamma-ray 
bursts are highly-collimated events. This supernova/gamma-ray burst
connection was dramatically confirmed when SN 2003dh was revealed 
in the afterglow of GRB 030329 (Stanek et al. 2003; Hjorth et al. 2003;
Kawabata et al. 2003).  

Here we summarize some of the background on spectropolarimetry of core-collapse 
supernovae and the evidence that they are generically asymmetric. We discuss the 
importance of the magneto-rotational instability for the collapse problem and some 
of the attendant physics that may be expected. We outline some of the important 
issues involved in doing MHD collapse and give a summary and critique of some 
recent attempts to merge MHD physics with core collapse physics.
 
\section {Results of Spectropolarimetry}

The first qualitative insight of the ``Texas" program of routine 
spectropolarimetry was that there is a distinct difference between 
Type Ia supernovae and core collapse events: Type II, Type IIn, Type IIb, Type 
Ib and Type Ic. The first systematic study (Wang et al. 1996) showed that core 
collapse supernovae are substantially polarized at the 1 \% level, but that Type 
Ia were generally substantially less polarized. As more data were added, it became 
clear that the polarization of the core-collapse supernovae was deeply intrinsic 
to the explosion mechanism. The polarization grows as the photosphere recedes into 
the ejecta and tends to be higher for events with less thick blanketing hydrogen 
envelopes (Wang et al. 1996, 2001; Leonard et al. 2001),  This implies that the 
basic machine that powers the explosion is asymmetric. Polarization of $\sim$ 1 \% 
implies an axis ratio of about 2 to 1 if interpreted in terms of ellipsoids of 
rotation (H\"oflich 1991). 

The data often show a well-defined orientation suggesting that the explosion was 
substantially bi-polar. Figure 1 shows the data for the Type II plateau event 
SN~1999em. The data fall on the same line in the Stokes parameter plane as a 
function of time and of wavelength.  This shows that there is a strongly favored 
axis to the geometry, hence that it is substantially bi-polar, a pattern repeated 
in several other events. We stress that there are exceptions. The Type IIb 
SN~1993J and the very similar event SN~1996cb showd large scatter in the Stokes 
parameter plane (Wang et al. 2001). 
SN~2002ap, a Type Ic that showed very high photospheric velocities in early 
phases revealed that the dominant axes associated with hydrogen, with oxygen, and 
with calcium were all oriented substantially differently (Wang et al. 2003b). 
Observational programs clearly have much more to teach us about the complexity of 
asymmetric supernovae and the physics involved in the asymmetry.

\begin{figure}
\begin{center}
  \includegraphics[height=.5\textheight]{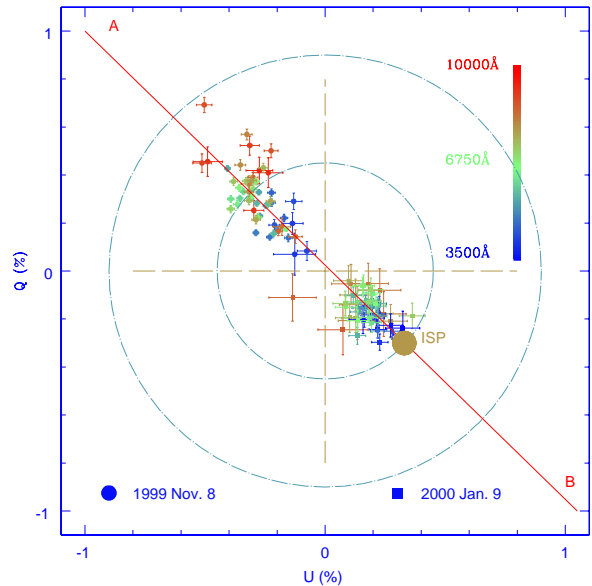}
\end{center}
  \caption{Two epochs of spectropolarimetry on the Type II plateau 
supernovae 1999em showing the bi-polar nature of the ejecta that falls 
along a single locus in the plane of the Stokes vectors as a function of 
time and wavelength (Wang et al. 2001)}
\end{figure}

\section{Asymmetric Core Collapse} 

We have learned that all core collapse supernovae are substantially asymmetric and 
often bi-polar. This alone does not prove that supernovae are exploded by jets, 
but numerical simulations (Khokhlov et al. 1999; Khokhlov \& H\"oflich 2001; 
H\"oflich, Wang \& Khokhlov 2001) have shown that bi-polar jets can, in principle, 
explode supernovae and produce these asymmetries with no aid from the classical 
powering process of neutrino deposition.  The origin of any such jets remains a 
mystery.  Rotation alone can induce asymmetric neutrino fluxes (Shimizu, Yamada, 
\& Sato 1994; Fryer \& Heger 2000), but rotation will inevitably lead to magnetic 
field amplification that can both produce MHD effects, including possibly jets 
(Wheeler et al. 2000, 2002; Akiyama et al. 2003), and affect neutrino transport 
(see \S 9 for a brief discussion).  In practice, neutrino transport, probably 
itself asymmetric and bi-polar will remain an important ingredient 
in the phenomenon. 

Asymmetries will also affect nucleosynthesis (Maeda 
et al. 2002; Nagataki et al. 2003). An important aspect of the jet-induced 
simulations is a characteristic feature of the chemical distribution. There will 
be a generic tendency for the iron-peak elements to be ejected along the jet 
direction with the traditional elements of bulk nucleosynthesis (oxygen, calcium) 
being ejected predominantly in the equatorial plane.  There is evidence that 
SN~1987A shows just that sort of configuration (Wang et al. 2002a). 
An interesting challenge to this picture is the recent data from a long CXO 
exposure of Cas A that clearly shows the jet and counter-jet structure long 
associated with Cas A, but predominantly in the element silicon, not, apparently, 
iron (Hwang et al. 2004). Another challenge is the displacement of the
central compact object to the south of the center of expansion of the
remnant, implying a ``kick" to the compact object of about 330 \kms\
roughly normal to the locus of the jet (Thorstensen, Fesen \& van den Bergh,
2001). It is possible that this complex dynamical structure is related to the 
multiple axes revealed in SN~2002ap (Wang et al. 2003b). 

The tendency for collapse explosions to be bi-polar suggests that at the very 
least rotation is involved to provide a special, well-defined axis.  There are 
strong arguments that rotation will naturally and unavoidably be attended with 
dynamo processes that generate and amplify magnetic fields.  It is probably 
inconsistent to consider rotation in either the collapse process or the stellar 
evolution that precedes it without simultaneously and self-consistently 
considering the attendant magnetic field. 

The ultimate problem of core collapse is one of three dimensions, rotation, 
magnetic fields, and neutrino transport.  We have suspected this all along, but 
the polarization of supernovae and jets from GRBs demands 
that the issue of substantial asymmetries be met head on.

\section{The Magneto-Rotational Instability and Core Collapse}

An important physical effect that must be considered in the context of core 
collapse is the magneto-rotational instability (MRI; Balbus \& Hawley 1991, 1998). 
Core collapse will lead to strong differential rotation near the
surface of the proto-neutron star even for initial solid-body
rotation of the iron core (Kotake, Yamada \& Sato 2003; Ott et al. 2003).
The criterion for instability to the MRI is a negative gradient
in angular velocity, as opposed to a negative gradient in
angular momentum for the Rayleigh dynamical instability. This
condition is generally satisfied at the surface of a newly formed
neutron star during core collapse and so the growth of magnetic
field by the action of the MRI is inevitable.  

More quantitatively,
when the magnetic field is small and/or the wavelength is long
(k $\mathrm{v_a}$ $<$ $\Omega$) the instability condition can be written
(Balbus \& Hawley 1991, 1998):
\begin{equation}
N^2+\frac{\partial\Omega^2}{\partial \mbox{ ln }\mathrm r} < 0,
\end{equation}
where N is the Brunt-V\"{a}is\"{a}l\"{a} frequency.  Convective
stability will tend to stabilize the MRI, and convective
instability to reinforce the MRI.  The saturation field given by
general considerations and simulations is approximately given by
the condition: $\mathrm{v_a}\sim$ $\lambda \Omega$ where
$\lambda \lta$ r or $\mathrm{B^2} \lta 4\pi \rho\mathrm{r^2\Omega^2}$
where $\mathrm{v_a}$ is the Alfv\'en velocity.

Akiyama et al.  (2003) have presented a proof-of-principle calculation
that the physics of the MRI is inevitable in the context of the differentially-
rotating environment of proto-neutron stars. The great power of the MRI to 
generate  magnetic field is that while it works on the rotation time scale of 
$\Omega^{-1}$ (as does simple field-line wrapping), the strength of the field 
grows exponentially.  This means that from a plausible seed field of $10^{10}$ to 
$10^{12}$ G that might result from field compression during collapse, only $\sim$ 
7 - 12 e-folds are necessary to grow to a field of $10^{15}$  G. Akiyama et al. 
(2003) have shown that for rotation that is at all times sub-Keplerian, this 
instability will naturally grow any seed field exponentially rapidly to a 
saturation level of order $10^{15}$ to $10^{16}$ G in a few 10s of milliseconds, a 
timescale longer than the initial bounce timescale, but much less than popular 
late-time neutrino-heating mechanisms that work over hundreds of milliseconds. 

Figure 2 shows the expected evolution of the angular velocity profile, the 
magnetic field and the associated MHD luminosity. The portion of the structure 
with decreasing angular velocity with radius, a generic feature at the boundary of 
the rotating proto-neutron star, represents structure that is unstable to 
the magneto-rotational instability. The predicted magnetic field is much 
larger than the quantum electrodynamic limit of $\sim 10^{13}$ G, but still 
smaller than the fields that would be directly dynamically important, of order 
$10^{17}$ to $10^{18}$ G. It remains to be seen whether this level of magnetic 
field will contribute substantially to asymmetries and jet formation in the 
explosions.  The effects on the equation of state are estimated to be negligible 
near the PNS where the density is high (Duan 2004; Akiyama et al. 2004), but
if a highly magnetic bubble is convected to a low density region, there
could be important effects. There could also be effects on the neutrino
cross sections as outlined briefly in \S 7. We note that these calculations
have not yet considered the de-leptonization phase when the neutron
star contracts and spins even faster, perhaps producing even
larger fields on timescales of seconds. 

\begin{figure}[htp]
\begin{center}
  \includegraphics[height=0.5\textheight]{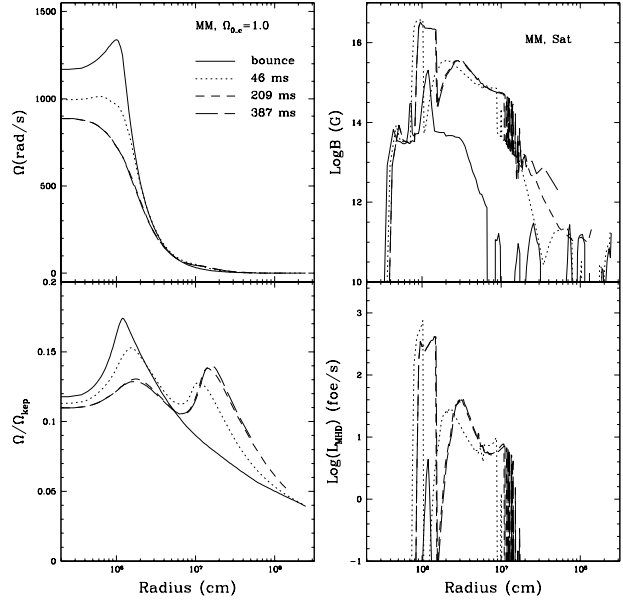}
\end{center}
  \caption{Angular  velocity, field strength and MHD luminosity(in units 
of $10^{51}$ erg s$^{-1}$) for a representative initial differential 
rotation of the iron core as a function of time from Akiyama et al. 
(2003)}
\end{figure}

The resulting
characteristic MHD luminosity (cf. Blandford \& Payne 1982) is:
\begin{equation}
\mathrm L_{\mathrm{MHD}}\sim \mathrm B^2 \mathrm r^3\Omega/2\sim 3\times
10^{52}\mbox{ erg } \mathrm s^{-1} \mathrm B_{16}^2 \mathrm
R_{\mathrm{NS.6}}^3\left (\frac{\mathrm
P_{\mathrm{NS}}}{10~\mbox{ms}}\right)^{-1}.
\end{equation}
If this power can last for a significant fraction of a second, a
supernova could result.
The energy of rotation is approximately
\begin{equation}
\mathrm E_{rot}\sim \mathrm 1/2 I_{NS}\Omega_{NS}^2 \sim 1.6\times
10^{50}\mbox{ erg } \mathrm M_{NS} \mathrm
R_{\mathrm{NS.6}}^2\left (\frac{\mathrm
P_{\mathrm{NS}}}{10~\mbox{ms}}\right
)^{-2}.
\end{equation}
A sufficiently fast rotation of the original iron core is needed to provide
ample rotation energy. This will also promote a strong MHD luminosity.

For collapse to form a black hole, the velocities will be
Keplerian and the associated, dynamo-driven, predominantly
toroidal field will have a saturation strength, B$^2\sim 4\pi \rho 
{\lambda^2} \Omega^2$ with $\lambda \lta$ r, of order B$\sim$10$^{16}$G
$\rho_{10}^{1/2}$ assuming motion, including the Alfv\'en speed,
near the speed of light near the Schwarzschild radius and a
characteristic density of order 10$^{10}$ g cm$^{-3}$
(MacFadyen \& Woosley 1999).  Fields this large could affect both
the dynamics  and the microphysics in the black hole-formation
problem.  Because of the nearly Keplerian motion in the black
hole case, the fields generated will be much closer to pressure
equipartition than in the neutron star case, and hence, perhaps,
even more likely to have a direct dynamical effect. The
associated MHD power in the black hole case would be roughly
$10^{52}-10^{53}$ erg s$^{-1}$.

The implication of this work is that the MRI is probably unavoidable in 
the differentially rotating ambience of core collapse for either 
``ordinary" supernovae and for those that produce gamma-ray bursts. 
Calculations that omit this physics are probably incorrect at some level. The 
magnetic field generated by the MRI should be included in any self-consistent 
calculation, but issues of how to capture this physics in numerical calculations 
are challenging. Balbus \& Hawley (1998) summarize their work showing that the 
specific outcome of MRI calculations depends on the initial field configuration. 
In 2D, an initial magnetic field aligned with the rotation axis will give a 
streaming instability, whereas a configuration with a finite RMS field but with 
zero mean field will give a chaotic, turbulent field. We return to this point 
below (\S 8).  

These implications need to be explored in much greater depth, but there is at 
least some possibility that the MRI may lead to strong MHD jets by the 
magneto-rotational (Meier, Koide \& Uchida 2001) or other mechanisms.  
A key point is that the relevant dynamics will be dictated by strong, 
predominantly toroidal fields that are generated internally, and are not 
necessarily the product of twisting of external field lines that is the basis 
for so much work on MHD jet and wind mechanisms.  Understanding the role of these 
internal toroidal fields in producing confining coronae (Hawley \& Balbus 2002) or 
jets (Williams, 2003), in providing the ultimate dipole field strength for 
both ordinary pulsars and magnetars (Duncan \& Thompson 1992), in 
setting the ``initial" pulsar spin rate after the supernova dissipates (that is, 
the ``final" spin rate from the supernova dynamicists point of view), and any 
connection to \grbs\ is in its infancy.

\section{Open Issues}

There are a large number of important open issues. Chief among them
are whether or not the rotation and magnetic fields associated with core
collapse lead to sufficiently strong MHD jets or other flow
patterns to explode supernovae. This issue touches on many others:
\begin{itemize}
\item Magnetic effects in the rotating progenitor star
\item Dynamos and saturation field strengths
\item Effect of large fields on the equation of state
\item Effect of large fields on the neutrino cross sections and transport
\item Effect of large fields on structure and evolution of the neutron star
\item Effect of large fields on jet formation
\item Relevance of MRI and field generation to GRBs and ``hypernovae"
\end{itemize}

\section{Dynamo Theory and Saturation Fields}

In traditional mean field dynamo theory, the turbulent velocity
field that drives the ``alpha" portion of the $\alpha - \Omega$
dynamo was specified and held fixed, but the turbulent velocity field cannot be
constant. The buildup of small scale magnetic field tends to inhibit
turbulence, cutting off the dynamo process for both small and
large scale fields. Since the small scale field tended to grow faster
than the large scale field, it appeared that the growth of the large
scale field would be suppressed (Kulsrud \& Anderson 1992;
Gruzinov \& Diamond 1994).  In these theories, the magnetic field energy
cascades to smaller length scales where it is ultimately dissipated
at the resistive scale.  Large scale fields tend to build up slowly,
if at all.

A proposed solution to this problem has been the recognition (Blackman \& Field
2000; Vishniac \& Cho 2001; Field \& Blackman 2002;
Blackman \& Brandenburg, 2002; Blackman \& Field 2002;
Kleeorin et al. 2002) that the magnetic helicity, {\bf H = A$\cdot$B}
is conserved in ideal MHD and that this conservation had not been
treated explicitly in mean field dynamo theory.  Incorporation of
this principle leads to an
``inverse cascade" of helical field energy to large scales that is
simultaneous with the cascade of helical field energy from the driving
scale to the dissipation scale.  Basically, the large scale helical field
and inverse cascade must exist with opposite magnetic helicity to that
of the field cascading to small scale.  The result (Blackman \&
Brandenburg 2002) is the rapid growth of large scale field in a kinematic
phase (prior to significant back-reaction) to a strength where the
field on both large and small scales is nearly in equipartition with
the turbulent energy density.  At that point, the back reaction sets
in and there tends to be a slower growth to saturation at field
strengths that can actually somewhat exceed the turbulent energy density.
It may be that the early, fast, kinematic growth is the only phase that
is important for astrophysical dynamos, especially in situations that
have open boundaries so that field can escape (Brandenburg, Blackman \&
Sarson 2003; Blackman \& Tan 2003) and that are very dynamic.
The collapse ambience is clearly one of those situations.

A related insight is that the rapid kinematic phase of field growth
can lead to magnetic helicity currents (Vishniac \& Cho 2001).
It is possible that these magnetic helicity currents can
transport power out of the system in twisting, propagating
magnetic fields. This is clearly reminiscent of jets or winds,
but the physics is rather different than any that has been
previously explored  in driving jets or winds.  This physics
needs to be explored in the context of supernovae and gamma-ray bursts.

Vishniac \& Cho (2001) argue that along with conservation of
magnetic helicity, {\bf H = A $\cdot$ B}, and the inverse
cascade of magnetic field energy to large scales, one will get a
current of magnetic helicity that can be crudely represented by
\begin{equation}
 J_{\mathrm H}\sim \mathrm B^2\lambda \mathrm v,
\end{equation}
where the characteristic length, $\lambda$, might be comparable to a
pressure scale height, $\ell_{\mathrm P}$ = (d ln P/dr)$^{-1}$, and
v $\sim \mathrm v_{\mathrm a}\sim \ell_{\mathrm P}\Omega$.  The energy
flux associated with this magnetic helicity current is J$_{\mathrm H}/
\lambda\sim\mathrm B^2\mathrm v_{\mathrm a}$, and so with $\mathrm
B^2 \sim \rho \ell_{\mathrm P}^2 \Omega^2$ the associated power is:
\begin{equation}
\mathrm L =\mathrm r^2\mathrm B^2\mathrm v_{\mathrm a}\sim
\mathrm B^2\mathrm r^2\ell_{\mathrm P}\Omega\sim\rho\mathrm r^5\Omega^3
\left(\frac{\ell_{\mathrm P}}{r}\right)^3.
\end{equation}
Note that the next-to-last expression on the RHS is essentially
just the characteristic Blandford-Payne luminosity; 
however, in this case the field is not externally given, but
provided by the dynamo process so that the final expression on
the RHS is given entirely in terms of local, internal
quantities.  The implication is that this amount of power is
available in an axial, helical field without twisting an
externally anchored field. Again, while this analysis has superficial
resemblance to other jet mechanisms, it involves rather
different physics and is self-contained.
Whether this truly provides a jet remains to be seen.  

Note that this process of creating a large scale field with an
MRI-driven dynamo with its promise of naturally driving axial,
helical flows does not require an equipartition field.  As pointed
out by Wheeler et al. (2002), the field does not have to have
equipartition strength and hence to be directly dynamically
important in order to be critical to the process of core
collapse.  The field only has to be significantly strong to
catalyze the conversion of the free energy of differential
rotation of the neutron star into jet energy.  As long as this
catalytic function is operative, the rotational energy should be
pumped into axial flow energy until there is no more
differential rotation.  For the case of stellar collapse, this
would seem to imply that, given enough rotational energy in the
neutron star,  this machine will work until there is a
successful explosion.  Even if the core collapses directly into
a black hole, or does so after some fall-back delay, the basic
physics outlined here, including magnetic helicity currents and
their associated power should also pertain to black hole formation.

\section{Neutrino Transport}

Fields of order 10$^{15}$ to 10$^{16}$ G that will characterize
both neutron star and black hole formation may affect neutrino
transport. With a large magnetic field, direct $\nu-\gamma$ interaction is
possible mediated by W and Z bosons.  This would allow
neutrino Cerenkov radiation, $\nu \rightarrow\nu +
\gamma$, and would enhance plasmon decay, $\gamma \rightarrow \nu + \nu$
(Konar 1997).

In addition, processes like $\nu \rightarrow\nu + \mathrm e^+ +\mathrm e^-$
would no longer be kinematically forbidden.  In that case, closed
magnetic flux loops can trap pairs.  The energy in pairs would
grow exponentially  to the point where annhilation cooling would
balance pair creation.  Thompson \& Duncan (1993) estimated that
an energy as much as E$_{\mathrm{pair}}\sim 10^{50}$ erg could be
trapped in this way.  This is not enough energy to cause a robust
explosion, but it is enough energy to drive the dynamics of core
collapse in a substantially different way, perhaps by inducing
anisotropic flow if the flux loops are themselves distributed
anisotropically.

With substantial magnetic fields, the cross section for inverse
beta decay, $\nu_{\mathrm e} +\mathrm n\rightarrow\mathrm p +
\mathrm e^-$, would become dependent  on neutrino momentum,
especially for asymmetric field distributions, which would be
the norm (Lai \& Qian 1998; Bhattacharya \& Pal 2003; Ando 2003;
Duan \& Qian 2004).

All these processes and more should be considered quantitatively
in core collapse to form neutron stars and black holes.

\section{Recent Work on Magnetic Core Collapse}

In this section we will review, compare and contrast some recent work on rotating 
magnetic collapse and related issues that pertain to understanding asymmetric core 
collapse. Other relevant work that we do not discuss in detail is in Burrows \& 
Hayes (1996) and Lai et al. (2001)
and references therein that discuss the effects 
of neutrino flux asymmetry. A crucial piece of information that can inform the 
discussion of potential MHD effects even in the absence of the explicit inclusion 
of magnetic fields is to give sufficient information from a rotating collapse to 
at least crudely estimate the saturation field according to the prescription $v_a 
\sim r \Omega$ or $B \sim \sqrt{4 \pi \rho} r \Omega$, that is, the angular 
velocity and the density profiles, or, even better, the product $\sqrt{\rho} \Omega$.  
Many studies of rotating collapse produce such 
information, but fail to present it explicitly. It would be very valuable if such 
information were presented explicitly as a function of time.

{\bf Akiyama et al. (2003)}  - As outlined above, Akiyama et al. did ``shellular" 
rotating collapse calculations with multi-group flux limited diffusion of 
neutrinos, no angular momentum transport (although that possibility was discussed) 
and a heuristic treatment of the MRI. They concluded that core collapse is 
generically susceptible to the MRI and that the MRI could be important.  They 
found fields of order $10^{15}$ to $10^{16}$ G could plausibly be generated in 
tens of milliseconds after bounce. Such fields are interestingly larger than the 
QED limit, but still not directly dynamically important. The magnetic field need 
not necessarily be dynamically important if the field can catalyze the dumping of 
the rotational energy of the neutron star into some useful, explosion-inducing 
form, jets or otherwise.  This basic energy requirement puts a premium on rapid 
rotation of the progenitor and the proto-neutron star in order to have a 
sufficiently large energy reservoir on which to draw. Of course, the rotational 
energy may be abetted by the large neutrino flux.

{\bf Thompson, Quataert, \& Burrows (2004)} - Thompson et al. also did ``shellular" 
rotating collapse calculations with a heuristic treatment of the MRI. Not 
surprisingly, given the similar assumptions and computations, they confirmed the 
field strength estimates of Akiyama et al. The new ingredient in this paper was to 
add viscous dissipation heating. Thompson et al. found that they could induce 
explosions for rapid enough rotation.

{\bf Fryer \& Warren (2004)} - In a series of works culminating, for now, in this 
paper, Fryer \& Heger (2000) and Fryer \& Warren (2002) have explored rotating core 
collapse. See also Fryer's contribution to these proceedings.  This work has used 
an SPH code with Fryer \& Heger and Fryer \& Warren examining the 2D case and 
Fryer \& Warren (2004) full 3D hydrodynamics. A feature that complicates the comparison 
of the results of this work with that from grid-based codes is that the SPH code 
yields prompt explosions in the basic non-rotating case, but no current grid-based 
code does so. The SPH code uses single energy, flux-limited diffusion. In the 
rotating calculations there are issues of angular momentum transport in SPH versus 
grid-based calculations. Fryer \& Heger (2000) and Fryer \& Warren (2002)
found that rotation alone could induce bi-polar, asymmetric explosions with axis ratios 
of 2 to 1, but the calculations were not run into the free-expansion phase, so it 
is not clear that this large asymmetry will survive as required by the 
spectropolarimetry. 

In their 3D calculation, Fryer \& Warren take note of 
significant evolution in the angular momentum distribution. An important factor is 
the tendency for low angular momentum matter to flow in along the rotation axis 
while larger angular momentum matter tends to halt along the equator.  This 
aspect of the dynamics cannot be captured in ``shellular" calculations, but 
should be manifested in 2D calculations.  Unfortunately, other papers have not 
commented specifically on this phenomenology which should be quite generic.  It 
would be useful in making comparisons if others were to do so. Fryer \& Warren 
(2004) do not include MHD, but use the heuristic prescription of the saturation 
field strength presented by Akiyama et al. to estimate the field strength for the 
angular velocity gradient and density they compute. Their estimates of the field 
strength are substantially less than found by Akiyama et al. even though, despite 
the very different dynamics (3D versus ``shellular"), the resulting angular 
velocity gradients are rather similar. The difference seems to be that, with a 
prompt explosion, the density declines rapidly, thus decreasing the implied 
Alfv\'en speed and hence saturation field for a given angular velocity gradient.

{\bf Buras, Rampp, Janka, \& Kifonidis (2003)} - Buras et al. do a sophisticated 
rotating collapse with Boltzmann neutrino transport on radial rays. These 
calculations have no magnetic effects, but are of the sort that can establish the 
ambience in which MHD effects will occur.  These calculations give bi-polar flow, 
but no explosion. The angular velocity profiles are not presented, so it is 
difficult to compare to other computations in that fundamental regard.

{\bf Ott, Burrows, Livne \& Walder (2003)} - Ott et al. used Livne's VULCAN/2D code 
to study rotating collapse.  They include no neutrino transport, but do present 
useful information on angular velocity profiles at certain epochs.  These 
calculations revealed the strong shear expected in core collapse and gave 
bi-polar flow patterns, but no explosion.

{\bf Kotake, Sawai, Yamada \& Sato (2004)} - Kotake et al. present 2D rotating, MHD 
collapse calculations using the ZEUS-2D code.  They incorporate an approximate 
neutrino cooling with a leakage scheme. They assume the initial field prior to 
collapse is predominantly toroidal and explore the effect on anisotropic neutrino 
radiation. They find more effective neutrino heating near the axis in a way that 
affects the dynamics. These calculations assume rapid pre-collapse 
rotation and pre-collapse magnetic fields in the range 
$5\times10^{9} - 10^{14}$ G. Such initial fields are probably 
unrealistically large. The calculations do produce phenomena that resemble MHD 
jets. The effects of field line wrapping are difficult to discriminate from the 
MRI, but Kotake et al. conclude that the MRI is likely to occur after 
bounce due to non-axisymmetric perturbations. 

{\bf Yamada \& Sawai (2004)} - Yamada \& Sawai also use ZEUS-2D but with a 
parametrized equation of state and no neutrinos. They assume rapid pre-collapse 
rotation and pre-collapse poloidal magnetic fields that are uniform, parallel 
to the rotation axis and with an amplitude of $\sim 10^{12}$ G. Again these large 
initial fields are probably unphysical. Yamada \& Sawai find large fields ``behind 
the shock" not in the core as for the pioneering calculation of LeBlanc \& Wilson 
(1971). Once again it is difficult to see whether the growth of field strength is 
due to field line wrapping, especially with the initial axial field, or some 
aspect of the MRI, or both.

{\bf Madokoro, Shimizu, \& Motizuki (2003)} - Madokoro et al. (see also Shimizu et 
al. 1994) explore non-rotating models in which a prolate, anisotropic neutrino 
radiation field is imposed. They find that such an anisotropic neutrino flux gives 
a larger explosion energy for given neutrino luminosity.

{\bf Ardeljan, Bisnovatyi-Kogan, Kosmachevskii \& Moiseenko (2004)} - Ardeljan et al. 
(see also Ardeljan, Bisnovatyi-Kogan \& Moiseenko 2004 and Moiseenko, Bisnovatyi-Kogan
\& Ardeljan 2004) present their own version of 2D MHD collapse and explosion for a 
collapsing bare white dwarf. They compute the collapse with rotation until the 
structure is nearly in hydrostatic equilibrium and then ``turn on" 
a field that is subsequently amplified. They explore both dipole and quadrupole 
initial fields.  The magnetic field subsequently grows to become comparable 
to the local pressure at which time an MHD shock is generated.  The formation of the 
MHD shock may be related to the low density associated with the bare white dwarf collapse.   
They get some mass ejection with an energy of about $5\times10^{50}$ ergs for a model in 
which the initial magnetic energy is a fraction $10^{-6}$ of the gravitational energy. 
If the gravitational energy corresponds to a neutron star with binding energy of order 
$10^{53}$ ergs this corresponds to an initial field of roughly $10^{15}$ G.  
Although the MRI is mentioned in Ardeljan, Bisnovatyi-Kogan \& Moiseenko and
by Moiseenko et al., few details are presented, so the mechanism of the field 
amplification is not clear.
Unique among the calculations summarized here, this work follows the neutron
star for $\sim$ 10 s as it contracts and speeds up. 

{\bf Hawley \& Balbus (2002)} - Hawley \& Balbus performed the first MHD simulation 
of a collapse-related environment in which the MRI and jet formation were explicit 
ingredients. They use the ZEUS algorithms to solve the MHD equations. This was a 
3D MHD simulation of the accretion of a torus of matter around a black hole. This 
is not the same as a true collapse calculation in the sense that there is no 
surrounding star, but it is still instructive. The torus accretes due to the 
turbulent stresses generated by the MRI. The resulting flow forms a hot, thick, 
nearly-Keplerian disk, a surrounding magnetized corona, and a jet up the axis. A 
key point is that their jet is not confined by hoop stress.  It is held out by the 
centrifugal barrier and held in by the pressure of the highly magnetic ($\beta << 
1$) corona. It is not clear how much this simulation would change if there were a 
surrounding, infalling star. Hawley \& Balbus note that there was no significant 
dynamical difference between simulations that included or omitted resistive 
heating. Hawley \& Balbus suggest that they get larger fields than in their closed 
box simulations, but do not discuss the reasons in any detail. They also do not 
explicitly discuss whether their fields are turbulent, changing sign on turbulent 
time scales, or are well-ordered and large scale.

{\bf Proga,  MacFadyen, Armitage, \& Begelman (2003)} - Proga et al. also explored 
a somewhat different problem, the rotating magnetic core collapse onto a black 
hole rather than onto a proto-neutron star, but the similarities and differences 
are again instructive. Proga et al. also employ ZEUS-2D MHD to do calculations of 
accretion of a helium envelope around a pre-existing black hole. They adopt a 
particular initial angular momentum distribution that does not necessarily 
correspond to the other collapse models discussed here. They adopt an initial 
field that is purely radial, again a configuration unlike any of the other work 
mentioned here. They find that a thick, Keplerian torus forms with the subsequent 
development of a Poynting-flux dominated jet. They argue that the MRI is active in 
producing amplification of their initial field, but give few details of the 
operation of the MRI in their models.  It is of particular interest that Proga et 
al. find their jet to be Poynting-flux dominated, since none of the canonical 
neutron star formation calculations seem to do so, even with rather large initial 
rotation and magnetic fields. This may be an artifact of their doing a calculation 
with a previously existing black hole onto which the helium envelope is dumped 
``impulsively." This may lead to a more evacuated, low density environment within 
the collapsing core and hence perhaps a tendency to be more nearly field dominated 
at a given field strength than if the collapse were followed {\it ab initio}
from an iron core.  
There may also be some differences in the way field amplification evolves in this 
environment. There could also be differences in the intrinsic environment of 
proto-neutron stars versus black holes. It would be very interesting to explore in 
more detail the differences between the calculation of Proga et al. and that of, 
for instance, Kotake et al. (2004).

This summary raises a central practical issue. The grand goal is to do a full 3D, 
sufficiently resolved (whatever that means) simulation of rotational magnetic core 
collapse with anisotropic neutrino transport.  Various groups, as just summarized, 
are taking the first steps in this direction.  To what degree can current (or 
future) simulations be regarded as adequate to capture the physics?  A case in 
point are simulations that may or may not reproduce the physics of the MRI. 

A basic issue here is well-illustrated by Balbus \& Hawley (1998) who point out 
that the Cowling dynamo theorem pertains to 2D MHD calculations. This theorem 
states that an axially-symmetric magnetic field cannot be produced by
fluid flow (Cowling 1957). The implication, as pointed out by Balbus \& Hawley 
(1998), is that sustained magnetic field amplification by axisymmetric turbulence is 
impossible in an isolated dissipative 2D system. Balbus \& Hawley show, as pointed 
out earlier, that the behavior of MRI simulations with initial vertical field and 
those with initial turbulent field ($<$B$>$ = 0) are very different with the 
former yielding channel flow solutions and the latter turbulent solutions.  In 2D 
simulations, however, both configurations
yield a decaying field after an initial transient 
growth phase. Three-dimensional simulations, on the other hand, show a rapid 
growth to a saturation field. Balbus \& Hawley argue that the results still retain 
some memory of the initial field configuration, but that the MRI works to grow the 
field exponentially rapidly even with $<$B$>$ = 0 as long as the simulation is 
adequately resolved. The time to reach saturation depends on the initial value of 
$<$B$^{2}$$>^{1/2}$, but the saturation level is basically independent of the 
value of the initial mean field.

What, then, are we to make of the various simulations described above, especially 
those that are in 2D and may or may not see the effects of the MRI?  It is 
possible that they are seeing the transient growth before the Cowling anti-dynamo 
effects set in.  Is that adequate for a dynamical situation like a supernova where 
the field only needs to act long enough to produce an explosion? Perhaps, but that 
does not seem satisfactory when the goal is to get the physics right. In addition, 
the 2D simulations reproduced in Balbus \& Hawley (1998) seem to show smaller 
ratios of peak fields to initial fields than the 3D calculation and that the field 
in 2D simulations decays on a timescale comparable to the rotation time, so there 
are indications already that one should not trust any 2D simulation to properly 
capture the physics of the MRI.  We have also seen that various groups have made 
radically different assumptions about the initial field structure -- axial, 
toroidal, radial -- when the simulations indicate that the initial field 
configuration may affect the results, especially in 2D. There has been no 
systematic study of this possibility. 

Another point to ponder is the stricture in the Cowling dynamo theorem that 
the system to which the theorem applies should be isolated and dissipative.  A 
supernova is dissipative, but it is not isolated; the whole point is to eject 
matter through previously existing boundaries.  The possibility of magnetic 
helicity currents as described by Vishniac \& Cho (2001; \S 6) may also be relevant in 
this context. Is it possible that MHD calculations that produce jets and outflow 
are thereby not ``isolated," and not subject to the anti-dynamo restrictions, and 
hence in some sense more trustworthy? Again this point has not been discussed at 
all in the literature.

This is not to deny that all the calculations described above have merit as this 
difficult regime is explored.  The lesson is clear, however that the MRI is tricky 
to simulate and care must be taken in computing, presenting, and comparing results 
among various simulations. 

\section{Conclusions}

Spectropolarimetric studies have shown that all core-collapse supernovae 
yet observed are significantly asymmetric, with geometries that can be 
predominantly bi-polar, but can also be more complex in ways that have yet to be 
sufficiently characterized or understood (see Wang et al. 2001, Wang et al. 
2003b). This data, along with direct imaging of objects like SN~1987A, the Crab 
nebula and Cas A, means that the dynamics and the radiative transfer, both photons 
and neutrinos, are very likely to be significantly asymmetric.  Account of this 
asymmetry must be made in the analysis of these events, including the derivation 
of such basic quantities as the ejecta mass and energy.

Core collapse is an intrinsically shearing environment and hence generically 
subject to dynamo-like instabilities such as the magneto-rotational instability. 
This means that both rotation and magnetic fields are intrinsic to the process of 
core collapse. This applies both to supernovae and to gamma-ray bursts, to the 
formation of neutron stars and of black holes.

Many of the points of this discussion have been made with the implicit assumption 
that the cause of the asymmetry of core collapse supernovae as revealed by the 
spectropolarimetry is in some way related to rotation which must be accompanied by 
some dynamo action and field growth. A last point concerns alternatives to this 
tacit assumption. Blondin, Mezzacappa, \& DeMarino (2003) have shown that standing 
spherical accretion shocks such as those that arise in non-rotating, non-magnetic 
core-collapse supernova models are unstable to $\ell$ = 1 and 2 modes of oscillation.
See also Blondin in these proceedings, Foglizzo (2002), and the calculations of 
full collapse with neutrino transport illustrating this instability described by 
Janka (2004) and by Scheck et al. (2004). This instability and the rapid 
growth of turbulence behind the shock is driven by the injection of vorticity as 
the shock is perturbed from spherical. This interesting result means that one 
must, at least, be cautious about interpreting bi-polar symmetry as the result of 
rotation.  On the other hand, sufficient rotation (never mind magnetic fields)  
may damp this particular instability (but see Scheck et al.). 
It is perhaps worth noting that this instability can only produce 
bi-polar structure whereas the observations show that strong deviations from this 
simple geometry are not uncommon. Of course, the simplest jet-induced models are 
also intrinsically axially symmetric and fail the same test. Clearly, this 
instability must be added to the zoo of possible phenomenology and treated 
self-consistently in future work.

The bottom line is that many new vistas have been opened by considering the 
intrinsic asymmetry of core-collapse supernovae and the ability of rotating, 
magnetic collapse to account for the observations. There is much work to be done 
to understand all the attendant physics.

\section*{Acknowledgments}
The authors are grateful to Tony Mezzacappa for his invitation and encouragement 
to attend this workshop, to the Institute for Nuclear Theory for providing a rich 
and fruitful environment, and to John Hawley and Chris Fryer for especially 
rewarding conversations at the meeting. They also thank Dave Meier and Peter 
Williams for valuable on-going discussion of these topics. JCW is deeply grateful 
for the contributions of his collaborators Lifan Wang, Peter H\"oflich and 
Dietrich Baade and the excellent staff at the VLT, all of whom were critical to 
the success of our spectropolarimetry program. This work was supported in part by 
NASA Grants NAG5-10766 and NAG5-9302 and by NSF Grant AST-0098644.



\begin{thebibliography}{0}


\bibitem{}
Ando, S. 2003, Phy. Rev. D, 68, 63002

\bibitem{}
Akiyama, S. Wheeler, J. C., Meier, D. \& Lichtenstadt, I, ApJ, 584, 954

\bibitem{aki04} 
Akiyama, S., Wheeler, J. C., Meier, D. L., \& Duncan, R. C. 2004, 
in Cosmic Explosions in Three Dimensions, eds. P. H\"{o}flich, P. Kumar, 
\& J. C. Wheeler (Cambridge: Cambridge University Press), in press

\bibitem{}
Ardeljan, N. V. Bisnovatyi-Kogan, G. S., \& Moiseenko, S. G. 2004, MNRAS, 
in press (astro-ph/0410234)

\bibitem{}
Ardeljan, N. V. Bisnovatyi-Kogan, G. S., Kosmachevskii, K. V. \& Moiseenko, S. G. 
2004, Astronomy, 47, 37. 

\bibitem{}
Balbus, S. A. \& Hawley, J. F.  1991, ApJ 376, 214

\bibitem{}
Balbus, S. A. \& Hawley, J. F. 1998, Review of Modern Physics, 70, 1

\bibitem{}
Bhattacharya, K. \& Pal, P. B. 2003, hep-ph/0209053

\bibitem{Blackman1}
Blackman E. G. \& Brandenburg, A. 2002, \apj, 579, 359

\bibitem{Blackman2}
Blackman, E. G. \& Field, G. B. 2000, \apj, 534, 984

\bibitem{Blackman3}
Blackman, E. G. \& Field, G. B. 2002, Phys. Rev. Lett., 89, 265007

\bibitem{Blackman4}
Blackman E. G. \& Tan, J. 2003, in ``Proceedings of the
International Workshop on Magnetic Fields and Star Formation: 
Theory vs. Observation," in press (astro-ph/0306424)

\bibitem{}
Blandford, R. D. \& Payne, D. G. 1982, MNRAS, 199, 833

\bibitem{2003ApJ...584..971B} 
Blondin, J.~M., Mezzacappa, A., \& DeMarino, C.\ 2003, \apj, 584, 971

\bibitem{Brandenburg}
Brandenburg, A., Blackman E. G. \& Sarson, G. R. 2003,
Adv. Space Sci., 32, 1835 

\bibitem{2003PhRvL..90x1101B} 
Buras, R., Rampp, M., Janka, H.-T., \& Kifonidis, K.\ 2003, Phys. Rev. 
Lett., 90, 241101

\bibitem{burrows96} 
Burrows, A., \& Hayes, J., 1996, Phys. Rev. Lett., 76, 352

\bibitem{}
Cowling, T. G. 1957, Magnetohydrodynamics, (New York: Interscience) 

\bibitem{dua04} 
Duan, H., \& Qian, Y. -Z. 2004, Phys. Rev. D, 69, 123004

\bibitem{}
Duan, H. 2004, private communication

\bibitem{duncan}
Duncan, R. C. \& Thompson, C. 1992, \apj, 392, L9

\bibitem{Fesen}
Fesen, R. A. 2001, ApJ Supp. , 133, 161

\bibitem{field}
Field, G. B. \& Blackman E. G. 2002, \apj, 572, 68

\bibitem{2002A&A...392..353F} 
Foglizzo, T.\ 2002, \aap, 392, 353

\bibitem{}
Fryer, C. L. \& Heger, A. 2000, ApJ, 541, 1033

\bibitem{2002ApJ...574L..65F} Fryer, C.~L.~\& 
Warren, M.~S.\ 2002, \apjl, 574, L65

\bibitem{2004ApJ...601..391F} Fryer, C.~L.~\& 
Warren, M.~S.\ 2004, \apj, 601, 391

\bibitem{Gruzinov}
Gruzinov, A. V. \& Diamond, P. H. 1994, Phys. Rev. Lett., 72, 1651

\bibitem{2002ApJ...573..738H} Hawley, J.~F.~\& 
Balbus, S.~A.\ 2002, \apj, 573, 738

\bibitem{}
Hjorth, J. et al. 2003, Nature, 423, 847

\bibitem{}
H\"oflich 1991, Astron \& Astrophys.,, 246, 481;


\bibitem{}
H\"oflich, P., Khokhlov, A. \& Wang, L. 2001, in Proc. of the 20th
Texas Symposium on Relativistic Astrophysics, eds. J. C. Wheeler \&
H. Martel, (New York: AIP), 459

\bibitem{}
Howell, D.~A., H{\" o}flich, P., Wang, L., \& Wheeler, J.~C.\ 2001, ApJ,
556, 302

\bibitem{}
Hwang, U. et al. 2004, ApJ Lett., 615, L117 


\bibitem{}
Kawabata, et al. 2003, ApJ, 593, L19


\bibitem{}
Khokhlov, A. \&  H\"oflich, P. 2001, in Explosive Phenomena
in Astrophysical Compact Objects, eds. H.-Y, Chang, C.-H. Lee
\& M. Rho, AIP Conf. Proc. No. 556, (New York: AIP), p. 301

\bibitem{}
Khokhlov  A.M., H\"oflich P. A., Oran E. S., Wheeler J.C.
Wang, L, \& Chtchelkanova, A. Yu. 1999, ApJ, 524, L107

\bibitem{Kleorin}
Kleeorin, N. I., Moss, D., Rogachevskii, I. \& Sokoloff, D. 2002,
\aap, \textbf{387}, 453

\bibitem{}
Konar, S. 1997, PhD. Thesis

\bibitem{2003ApJ...595..304K} 
Kotake, K., Yamada, S., \& Sato, K.\ 2003, \apj, 595, 304 

\bibitem{2004ApJ...608..391K} 
Kotake, K., Sawai, H., Yamada, S., \& Sato, K.\ 2004, \apj, 608, 391 

\bibitem{kulsrud}
Kulsrud, R. M. \& Anderson, S. W. 1992, ApJ, 396, 606

\bibitem{}
Lai, D. \& Qian, Y.-Z. 1998, ApJ, 505, 844

\bibitem{lai01} Lai, D., Chernoff, D. F. \& Cordes, J.
   M. 2001, \apj, 549, 1111


\bibitem{}
Leonard, D.~C., Filippenko, A.~V., Barth, A.~J.,
\& Matheson, T.\ 2000, ApJ, 536, 239

\bibitem{}
Leonard, D.~C.~\& Filippenko, A.~V.\ 2001, PASP, 113, 920

\bibitem{}
Leonard, D.~C., Filippenko, A.~V., Ardila, D.~R.,
\& Brotherton, M.~S.\ 2001, ApJ, 553, 861

\bibitem{}
Leonard, D.~C., Filippenko, A.~V., Chornock, R. \& Foley, R. J.\ 2002,
PASP, 114, 1333 

\bibitem{macfayden}
MacFadyen, A. \& Woosley, S. E. 1999, \apj,  \textbf{524}, 262

\bibitem{}
Maeda, K., Nakamura, T., Nomoto, K., Mazzali, P., Patat, F.
\& Hachisu, I. 2002, ApJ, 565, 405

\bibitem{2003ApJ...592.1035M} 
Madokoro, H., Shimizu, T., \& Motizuki, Y.\ 2003, \apj, 592, 1035

\bibitem{2000ApJS..128..615M}
Marietta, E., Burrows, A., \& Fryxell, B.\ 2000, ApJS, 128, 615

\bibitem{meier}
Meier, D. L., Koide, S. \& Uchida, Y. 2001, Science, 291, 84

\bibitem{}
Moiseenko, S. G., Bisnovatyi-Kogan, G. S. \& Ardeljan, N. V. 2004,
in ``1604-2004 Supernovae as Cosmological Lighthouses,"
eds. M. Turatto et al. (San Francisco: ASP) astro-ph/0410330 

\bibitem{}
Nagataki, S., Mizuta, A., Yamada, H., Takabe, H. \& Sato, K. 2003,
ApJ, ApJ, 596, 401

\bibitem{ott04} Ott, C. D., Burrows, A., Livne, E., \&
Walder, R. 2004, \apj, 600, 834


\bibitem{2003ApJ...599L...5P} 
Proga, D., MacFadyen, A.~I., Armitage, P.~J., \& Begelman, M.~C.\ 
2003, \apjl, 599, L5


\bibitem{}
Scheck, L., Plewa, T., Janka, H. Th., Kifonidis, K. \& M\"uller, E.
2004, Phys, Rev. Lett., 92, 011103-1

\bibitem{}
Shimizu, T., Yamada, S., \& Sato, K. 1994, ApJ. Lett., 432, L119

\bibitem{stanek}
Stanek, K. Z. et al. 2003, \apj, 591, L17

\bibitem{thompson}
Thompson, C.~\& Duncan, R.~C.\ 1993, \apj, 408, 194

\bibitem{tho04b} 
Thompson, T. A., Quataert, E., \& Burrows, A. 2004, preprint (astro-ph/0403224)

\bibitem{}
Thorstensen, J. R., Fesen, R. A. \& van den Bergh, S. 2001, ApJ, 122, 297

\bibitem{vishniac}
Vishniac, E. T. \& Cho, J. 2001, \apj, \textbf{550}, 752

\bibitem{}
Wang, L., Baade, D., H\"oflich, P. \& Wheeler, J. C. 2002b, ESO
Messenger, No. 109, 47

\bibitem{}
Wang, L., Howell, D.~A., H{\" o}flich, P., \& Wheeler, J.~C.\ 2001, ApJ,
550, 1030 

\bibitem{}
Wang, L., Wheeler, J.~C., \& H\"oflich, P.\ 1997, ApJ. Lett., 476, L27

\bibitem{}
Wang, L., Wheeler, J.~C., Li, Z., \& Clocchiatti, A.\ 1996, ApJ, 467,
435

\bibitem{}
Wang, L. et al.\ 2002a, ApJ, 579, 671 

\bibitem{}
Wang, L. et al.\ 2003a, ApJ, 591, 1110   

\bibitem{}
Wang, L. et al.\ 2003b, ApJ, 592, 457   

\bibitem{2004ApJ...604L..53W} 
Wang, L., Baade, D., H{\"o}flich, P., Wheeler, J.~C., Kawabata, K., \& Nomoto, K.\ 
2004, \apjl, 604, L53 

\bibitem{}
Wheeler, J. C. 2004, in Cosmic Explosions in Three Dimensions:
Asymmetries in Supernovae and Gamma-Ray Bursts, eds P. H\"oflich,
P. Kumar \& J. C. Wheeler (Cambridge: Cambridge University Press),
astro-ph/0401323

\bibitem{}
Wheeler, J.~C., Meier, D.~L. \& Wilson, J.~R.\ 2002, ApJ, 568, 807

\bibitem{}
Wheeler, J. C., Yi, I., H\"oflich, P. \& Wang, L. 2000, ApJ, 537, 810

\bibitem{williams}
Williams, P. T., 2003, IAOC Workshop ``\textit{Galactic Star
Formation Across the
Stellar Mass Spectrum," ASP Conference Series}," ed. J. M. De Buizer,
in press (astro-ph/0206230)

\bibitem{2004ApJ...608..907Y} Yamada, S.~\& Sawai, 
H.\ 2004, \apj, 608, 907 


\end{thebibliography}


\end{document}